\documentclass[3p,times,twocolumn]{elsarticle}
 \biboptions{comma,sort&compress}

\usepackage{graphicx}
\usepackage{here}
\usepackage{ecrc}

\usepackage{dcolumn}
\usepackage{bm}
\usepackage{slashed}
\usepackage{multirow}
\usepackage{color}
\usepackage{ulem}
 \usepackage{epstopdf}
\volume{00}

\firstpage{1}

\journalname{Nuclear and Particle Physics Proceedings}

\runauth{}

\jid{nppp}

\jnltitlelogo{Nuclear and Particle Physics Proceedings}

\usepackage{amsmath}
\usepackage{amssymb}

\usepackage[figuresright]{rotating}

\begin{document}

\begin{frontmatter}

\title{ The Mass and Decay Properties of the $1^{-+}$ Light Hybrid Meson}
 \cortext[cor0]{Talk given at 20th International Conference in Quantum Chromodynamics (QCD 17),  3 July - 7 July 2017, Montpellier - FR}
 \author[label1]{Zhuo-Ran Huang}
\ead{huangzhuoran@126.com}
\address[label1]{Zhejiang Institute of Modern Physics, Zhejiang University, Zhejiang Province, 310027, P. R. China}
 \author[label1]{Hong-Ying Jin\fnref{fn1}}
    \fntext[fn1]{Speaker, Corresponding author.}
\ead{jinhongying@zju.edu.cn}
 \author[label2]{T.G. Steele}
    \ead{tom.steele@usask.ca}
\address[label2]{Department of Physics and Engineering Physics, University of Saskatchewan, Saskatoon, Saskatchewan, S7N 5E2, Canada}
 \author[label3]{Zhu-Feng Zhang}
\ead{zhangzhufeng@nbu.edu.cn}
\address[label3]{Physics Department, Ningbo University,
Ningbo, 315211, P.R. China}
\pagestyle{myheadings}
\markright{ }
\begin{abstract}

We calculate the complete form of the dimension-8 condensate contributions
in the  two-point correlator of the ($1^{-+}$,$0^{++}$) light hybrid current considering the operator mixing under renormalization. We find the inclusion these higher power corrections as well as the update of $\langle g^3G^3\rangle$ increase the QCD sum rule mass prediction for the $1^{-+}$ light hybrid. The obtained conservative mass range 1.72--2.60 GeV does not favor the $\pi_1(1400)$ and the $\pi_1(1600)$ to be pure hybrid states and suggests the $\pi_1(2015)$ observed by E852 is more likely to have much of a hybrid constituent. We also study the $b_1\pi$ and $\rho\pi$ decay patterns of the $1^{-+}$ light hybrid with light-cone QCD sum rules. We obtain a relatively large partial decay width of the $b_1\pi$ mode, which is consistent with the predictions from the flux tube models and lattice QCD. More interestingly, using the tensor interpolating current we find the partial decay width of the $\rho\pi$ mode is small due to the absence of the leading twist contribution in the light-cone expansion of the correlation function.
\end{abstract}

\end{frontmatter}
\section{Introduction}

The hybrid mesons with an exotic quantum number $1^{-+}$ are expected to be contained in one of the lowest-lying hybrid nonets. Therefore the identification of the $1^{-+}$ light hybrid state has long and continuously been an intriguing topic in hadronic physics. There are three candidates with $J^{PC}=1^{-+}$ observed in the experiments: $\pi_1(1400)$, $\pi_1(1600)$ and $\pi_1(2015)$ \cite{Patrignani:2016xqp}. These non-$q\bar{q}$ states can be interpreted as hybrids, four-quark states,  molecule states or their mixing. In this talk, we summarize our works \cite{Huang:2014hya} on calculating the mass of the $1^{-+}$ light hybrid meson from QCD sum rules (QCDSR) and \cite{Huang:2016upt}
on studying the partial decay widths of the decay modes $\pi_1\rightarrow b_1\pi$ and $\pi_1\rightarrow \rho\pi$ within the framework of light-cone QCD sum rules (LCSR).

\section{Mass of the $1^{-+}$  light hybrid from QCDSR}

\subsection{QCD expression of the two-point correlation function}

The starting point of our QCDSR analysis is the two-point correlation function
\begin{eqnarray}
\Pi_{\mu\nu}(q^{2}) & = & i\int d^{4}xe^{iqx}\left\langle 0\left|T\left[j_{\mu}(x)j_{\nu}^{+}(0)\right]\right|0\right\rangle \label{eq:1}\\
 & = & (q_{\mu}q_{\nu}-q^{2}g_{\mu\nu})\Pi_{v}(q^{2})+q_{\mu}q_{\nu}\Pi_{s}(q^{2}),\nonumber
 \end{eqnarray}
where $j_{\mu}(x)=g\bar{q}(x)iG_{\mu\nu}(x)\gamma_{\nu}q{(x)}$, and
the invariants $\Pi_{v}(q^{2})$ and $\Pi_{s}(q^{2})$ are
respectively the contributions from the $1^{-+}$ and the $0^{++}$ states.

The correlation function obeys the dispersion relation
\begin{equation}
\Pi_{v/s}(q^{2})=\frac{1}{\pi}\int_{0}^{\infty}ds\frac{\textrm{Im}\Pi_{v/s}(s)}{s-q^{2}-i\epsilon},
\label{eq:20}
\end{equation}
which relates the correlation function calculated perturbatively using the operator product expansion (OPE) for the large Euclidian $q^2$ to the hadronic spectral function measured experimentally.

Different authors have made efforts on the study of the $1^{-+}$ light hybrid meson within the framework
of QCD sum rules in the past 35 years,. The Leading order (LO) results of the perturbative and the $d\leqq6$ condensate terms in the OPE of $\Pi_{v}(q^{2})$  were calculated by several groups and summarized in \cite{Narison:2009vj}.

The next-to-leading (NLO) radiative corrections to the perturbative terms, the dimension-4 $\langle \alpha_s G^2\rangle$ terms, and the dimension-6 $\alpha_{s}\left\langle \overline{q}q\right\rangle ^{2}$ terms were included in \cite{Chetyrkin:2000tj,Jin:2000ekandJin:2002rw}.

Moreover, the contributions due to the short distance tachyonic gluon mass and the instanton effects beyond the original SVZ expansion were respectively calculated in \cite{Chetyrkin:2000tj} and \cite{zhangthesis}.

In \cite{Huang:2014hya}, we calculated the $d=8$ condensate terms in the OPE. The bilinear $d=8$ quark operators mix with the gluon operators $\langle G^4 \rangle$ in the lowest order under renormalization. We calculated the corrections to the  $\langle G^4 \rangle$ terms due to this mixing and examined explicitly the cancellation of the mass singularities, as was done for the vector current \cite{Broadhurst:1985jsandGrozin:1994hd}.

In the numerical analysis, we omit the instanton effects which are much less important than the radiative corrections in the $1^{-+}$ channel as shown in \cite{zhangthesis}. We will not show the full results of the OPE here for simplicity, which can be found in  \cite{Huang:2014hya}.

\subsection{Numerical analysis of the $1^{-+}$ mass}

By using the well tested "single narrow resonance minimal duality ansatz", the spectral function can be parametrized as
\begin{equation}
\textrm{Im}\Pi_v^{\textrm{phen}}(s)=\pi f_H^2 m_H^4 \delta(s-m_H^2)+\textrm{Im}\Pi_v^{\textrm{OPE}}(s)\theta(s-s_0),
\end{equation}
where $m_H$ is the mass of the lowest-lying $1^{-+}$ hybrid, $s_0$ is the continuum threshold, and $f_H$ is the resonance's coupling to the current.

The above phenomenological model can be related to the QCD expression of the correlation function through the dispersion relation. After applying the Borel operator
\begin{equation}
\hat B\equiv \lim\limits_{\substack{Q^2,n\rightarrow \infty\\ n/Q^2=\tau}}
\frac{(Q^2)^{n}}{(n-1)!}\left(- \frac{\rm d}{{\rm d} Q^2}\right)^n,
\end{equation}
 which improves the convergence of the OPE series and also enhances the the ground state contribution in the integral of the spectral function, we get the mater equation of QCD sum rules:
 \begin{equation}
\label{eq:qcdsr}
\hat{B}\Pi_v^{\textrm{OPE}}(\tau)=\hat{B}\Pi_v^{\textrm{phen}}(\tau,s_0,f_H,m_H),
\end{equation}
where we have $\hat{B}\Pi_v^{\textrm{phen}}(\tau,s_0,f_H,m_H)=\frac{1}{\pi}\int_0^\infty{\rm Im}\Pi_v^{\textrm{phen}}(s)e^{-s\tau}ds$ through the dispersion relation.

Numerically, in order to obtain reliable predictions, we use the Monte-Carlo based weighted-least-square method \cite{Leinweber:1995fn} to match the two sides of the master equation \eqref{eq:qcdsr} in the sum rule window. In this method,
the continuum threshold $s_0$, which is (in principal) a free parameter in the original SVZ sum rules, can be rigorously constrained.

In the Monte-Carlo based matching procedure, the sum rule window is devided at $\tau_j=\tau_\textrm{min}+(\tau_\textrm{max}-\tau_\textrm{min})\times(j-1)/(n_B-1)$, where $\tau_\textrm{min}$ and $\tau_\textrm{max}$ are respectively the lower bound and upper bound of the sum rule window, and we set $n_B=21$ in our analysis. The phenomenological outputs, $m_H$, $f_H$ and $s_0$ can be obtained by
minimizing
\begin{equation}
\chi^2=\sum_{j=1}^{n_B}\frac{(\Pi^{\textrm{OPE}}(\tau_j)-\Pi^{\textrm{phen}}(\tau_j,s_0,f_H,m_H))^2}{\sigma_{\textrm{OPE}}^2(\tau_j)},
\end{equation}
where $\sigma_{\textrm{OPE}}(\tau_j)$ is the standard deviation of $\Pi_v^{\textrm{OPE}}(\tau_j)$, estimated by randomly generating 200 sets of Gaussian distributed phenomenological inputs with 10\% uncertainties.

\begin{table*}[hbt]
\setlength{\tabcolsep}{1.5pc}
\newlength{\digitwidth} \settowidth{\digitwidth}{\rm 0}
\catcode`?=\active \def?{\kern\digitwidth}
\caption{Different input phenomenological parameters (at scale $\mu_0=1$\,GeV).}
\label{tab:input}
\begin{tabular*}{\textwidth}{@{}l@{\extracolsep{\fill}}cccc}
\hline
 & $\Lambda_{\textrm{QCD}}/\textrm{GeV}$ & $\langle \alpha_s G^2\rangle/\textrm{GeV}^4$ &  $m_q/\textrm{GeV}$ \\
\hline
Set I & $0.353$ & $0.07$& $0.007$ \\
\hline
Set II & $0.353$ & $0.07$& $0.007$\\
\hline
 & $\langle g^3 G^3\rangle$ & $\alpha_s\langle\bar qq\rangle^2/\textrm{GeV}^4$ & $\langle g\bar qGq\rangle$\\
 \hline
Set I & $8.2\,\textrm{GeV}^2 \langle \alpha_s G^2\rangle$ & $1.5\times10^{-4}$ & $0.8\,\textrm{GeV}^2 \langle \bar qq\rangle$\\
\hline
Set II & $1.2\,\textrm{GeV}^2 \langle \alpha_s G^2\rangle$ & $1.5\times10^{-4}$ & $0.8\,\textrm{GeV}^2 \langle \bar qq\rangle$\\
\hline
\end{tabular*}
\end{table*}

The central values of the QCD parameters used in our analysis are listed in Table~\ref{tab:input}. Set I are from a recent review article of QCD sum rules \cite{Narison:2014wqa}.  We use different values of $g^3\langle G^3\rangle$ in set I and set II, which are respectively estimated from (I) Charmonium sum rules \cite{Narison:2011rn} and (II) dilute gas instantons and lattice calculations \cite{Novikov:1981xiandD'Elia:1997ne}). The latter is the one used in previous sum rule analysis \cite{Jin:2000ekandJin:2002rw,Narison:2009vj,Zhang:2013rya}. We consider the violation of vacuum saturation in estimating the dimension-6 (up to 3) and 8 (up to 5) condensates.

By generating 2000 sets of Gaussian distributed inputs with 10\% uncertainties (of which the central values are the ones in set I and set II), for each set we obtain a set of phenomenological outputs $s_0$, $f_H$ and $m_H$ by minimizing $\chi^2$. After this procedure we can estimate the central values and uncertainties of the outputs.

Considering possible violation of factorization, different values of $\langle g^3G^3\rangle$, we obtain a quite conservative mass range 1.72--2.60\,GeV. Given that we have taken into account all effects that can influence the sum rule mass prediction considerably, this range strongly suggest that the $\pi_1(1400)$ and the $\pi(1600)$ may not be pure hybrid states. Only the mass of the unconfirmed $\pi_1(2015)$ is covered by the mass range, suggesting further experimental study on this state is important.

\section{Partial decay widths of $\pi_1\rightarrow b_1\pi$ and $\pi_1\rightarrow \rho\pi$}
\subsection{Formalism of the light-cone expansion}
In the frame work of light-cone QCD sum rules, one considers a current-current correlation function that involves an on-shell state in its matrix element. Our study of  the decay modes $\pi_1\rightarrow b_1\pi$ and $\pi_1\rightarrow \rho\pi$ begin with the following correlation function:
\begin{equation}
\label{eq:master1}
\Pi^{T,D}(k,p)=i\int d^4 xe^{ik\cdot x}\langle \pi(q)|T\{J^{T,D}(x)J^{H^\dagger}(0)\}|0\rangle,
\end{equation}
where $p$, $k$ and $q$ are respectively the momenta of $\pi_1$, $b_1$ or $\rho$ and $\pi$, which satisfy the four-momentum conservation $p=k+q$. We use
$J^{H}_\mu=\frac{\sqrt2}{2}(\bar{u}G_{\mu\nu}\gamma_{\nu}u-
\bar{d}G_{\mu\nu}\gamma_{\nu}d)$, $J^T_{\mu\nu}=\bar{d}
\sigma_{\mu\nu}u$ and $J^D_\mu=\bar{d}\overleftrightarrow{D}_\mu\gamma_5u$
to study the partial decay widths of the decay modes
$\pi_1^0\rightarrow b_1^+\pi^-$ and $\pi_1^0\rightarrow \rho^+\pi^-$,
of which the results are the same as those for $\pi_1^0\rightarrow b_1^-\pi^+$ and
$\pi_1^0\rightarrow \rho^-\pi^+$.

The correlation function satisfies the double dispersion relation, which connects the hadronic decay amplitudes to the light-cone expansion. The double dispersion relation reads
\begin{eqnarray}\label{dispersionrelation}
\label{eq:7}
\begin{split}
&\Pi(k^2,p^2)\\
=&\int_0^\infty ds_1 \int_0^\infty ds_2 \frac{\rho(s_1,s_2)}{(s_1-k^2-i\epsilon)(s_2-p^2-i\epsilon)}\\
&+\textrm{subtractions},
\end{split}
\end{eqnarray}
where the subtractions eliminate the infinities from the dispersion integral.

After taking the Borel transformation twice respectively with respect to $p^2$ and $k^2$,
the subtraction terms can be removed and we get the exponential form the light-cone sum rules:
\begin{eqnarray}
\label{eq:8}
\begin{split}
&\mathcal{B}_{k^2}^{\frac{1}{\sigma_1}}\mathcal{B}_{p^2}^{\frac{1}{\sigma_2}}\Pi(k^2,p^2)\\
=&\int_0^{\infty} ds_1\int_0^\infty ds_2\;e^{-s_1\sigma_1}e^{-s_2\sigma_2}\;
\rho(s_1,s_2).
\end{split}
\end{eqnarray}

On the phenomenological side of the sum rules, the correlation function can be expressed through inserting intermediate hadronic states. The strong couplings we are interested in and the decay constants are defined through:

\begin{eqnarray}
\label{eq:2}
\begin{split}
\mathcal{M}(\pi_1\rightarrow\rho\pi)
=&ig_\rho\varepsilon_{\alpha\beta\rho\sigma}\epsilon^{*\alpha}\eta^\beta k^\rho p^\sigma,\\
\mathcal {M}(\pi_1\rightarrow b_1\pi)
=&ig_{b_1}^1(\eta\cdot\epsilon^*)+ig_{b_1}^2(\eta\cdot k)(\epsilon^*\cdot p),\\
\langle 0|J^{H}_\mu(0)|\pi_1\rangle=&f_{\pi_1}m_{\pi_1}^3\eta_\mu,\\
\langle0|J^T_{\mu\nu}(0)|b_1\rangle=&if^T_{b_1}\varepsilon_{\mu\nu\rho\sigma}\epsilon^\rho k^\sigma,\\
\langle0|J^T_{\mu\nu}(0)|\rho\rangle=&if^T_{\rho}(k_\mu\epsilon_\nu-k_\nu\epsilon_\mu),\\
\langle0|J^D_\mu(0)|b_1\rangle=&f_{b_1}\epsilon_\mu,
\end{split}
\end{eqnarray}
where $\epsilon_\mu$ and $\eta_\mu$ are polarization vectors.

The spectral density $\rho(k^2,p^2)$ can be obtained by taking another double Borel transformations on \eqref{eq:8}:
\begin{eqnarray}
\label{eq:9}
\rho(s_1,s_2)=\mathcal{B}_{-\sigma_1}^{\frac{1}{s_1}}\mathcal{B}_{-\sigma_2}^{\frac{1}{s_2}}\mathcal{B}_{k^2}^{\frac{1}{\sigma_1}}\mathcal{B}_{p^2}^{\frac{1}{\sigma_2}}\Pi(k^2,p^2).
\end{eqnarray}

On the QCD side, the correlation function can be expanded near the light cone. Equating the QCD expressions and phenomenological expressions and subtract the continuum contributions, we get the formulae to perform the numerical analysis
(here due to space limitations we only present the one for $g_{b_1}^1$ obtained
from using the tensor interpolating current):
\begin{eqnarray}
\label{eq:11}
\begin{split}
&f^{T}_{b_1}f_{\pi_1}m_{\pi_1}^3g_{b_1}^1 e^{-m_{b_1}^2\sigma_1-m_{\pi_1}^2\sigma_2}\\
=&\int_0^{s_{01}} ds_1\int_0^{s_{02}} ds_2\;e^{-s_1\sigma_1}e^{-s_2\sigma_2}\;\\
&\cdot
\mathcal{B}_{-\sigma_1}^{\frac{1}{s_1}}\mathcal{B}_{-\sigma_2}^{\frac{1}{s_2}}
\mathcal{B}_{k^2}^{\frac{1}{\sigma_1}}\mathcal{B}_{p^2}^{\frac{1}{\sigma_2}}\Pi_{b1;1}^T(k^2,p^2)\,,
\end{split}
\end{eqnarray}
where $s_{01}$ and $s_{02}$ are the continuum thresholds.

Again for simplicity we only show the results of the light-cone expansion corresponding to the tensor current. They are
\begin{eqnarray}
\begin{split}
&\mathcal{B}_{k^2}^{\frac{1}{\sigma_1}}\mathcal{B}_{p^2}^{\frac{1}{\sigma_2}}\Pi_{b1;1}^T(k^2,p^2)\\
=&-\frac{\sqrt{2}\pi f_\pi m_\pi^2}{108(m_u+m_d)}\langle \alpha_sG^2\rangle\biggl\lbrace\frac{1}{2}[\phi_\sigma'(u_0)-\phi_\sigma'(\bar{u}_0)]\\
&+3[\phi_p(u_0)+\phi_p(\bar{u}_0)]+3(\phi_p^{[u]}+\phi_p^{[\bar{u}]})\biggl\rbrace,
\end{split}
\end{eqnarray}
\begin{eqnarray}
\label{eq:21}
\begin{split}
&\mathcal{B}_{k^2}^{\frac{1}{\sigma_1}}\mathcal{B}_{p^2}^{\frac{1}{\sigma_2}}\Pi_{b1;2}^T(k^2,p^2)\\
=&-\frac{\sqrt{2}f_\pi m_\pi^2}{(m_u+m_d)}(\mathcal{T}^{[\alpha_1]}+\mathcal{T}^{[\alpha_2]})\frac{1}{\sigma}\\
&+\frac{\sqrt{2}\pi f_\pi m_\pi^2}{108(m_u+m_d)}\langle \alpha_sG^2\rangle\\
&\cdot\biggl\lbrace[\phi_\sigma(u_0)+\phi_\sigma(\bar{u}_0)](\sigma_1-\sigma_2)\\
&+6(\phi_p^{[u]}+\phi_p^{[\bar{u}]})\sigma_2\biggl\rbrace,
\end{split}
\end{eqnarray}
\begin{eqnarray}
\label{eq:22}
\begin{split}
&\mathcal{B}_{k^2}^{\frac{1}{\sigma_1}}\mathcal{B}_{p^2}^{\frac{1}{\sigma_2}}\Pi_{\rho}^T(k^2,p^2)\\
=&\frac{\sqrt{2}\pi f_\pi m_\pi^2}{108(m_u+m_d)}\langle \alpha_sG^2\rangle\biggl\lbrace[\phi_\sigma(u_0)+\phi_\sigma(\bar{u}_0)]\sigma\\
&-6(\phi_p^{[u]}+\phi_p^{[\bar{u}]})\sigma_2\biggl\rbrace,
\end{split}
\end{eqnarray}
where the Borel variable is $\sigma=\sigma_1+\sigma_2$, and the definitions of the notations can be found in \cite{Huang:2016upt}.

\subsection{Numerical analysis for $\pi_1\rightarrow b_1\pi$}

In our numerical analysis, we use the standard sum rule stability criteria, i.e. we obtain the optimal outputs by demanding
that they are insensitive to the variation of the external parameters, the Borel parameter $\sigma=\sigma_1+\sigma_2$ and the continuum thresholds $s_{01}$ and $s_{02}$. Since the mass of the $1^{-+}$ light hybrid is still uncertain, we consider three different values according to the experimental candidates, i.e., we use $m_{\pi_1}$ = 1.6\,GeV, 1.8\,GeV and 2.0\,GeV. For the other input parameters, numerically we adopt $f_{\pi_1}=0.025\,\textrm{GeV}$ \cite{Narison:2009vj,Huang:2014hya},
$m_{b_1}=1.235\,\textrm{GeV}$, $f^T_{b_1}(2~\textrm{GeV})=0.18\,\textrm{GeV}$ \cite{Jansen:2009yh} and $f_{b_1}(2~\textrm{GeV})=0.18\,\textrm{GeV}$ \cite{Reinders:1984sr}. We omit the detail of the lengthy numerical analysis here, which can be found in \cite{Huang:2016upt}. Instead we list the results in Table~\ref{tab:width}. The results in the third row are the optimal ones from the stability criteria, which are $\Gamma(\pi_1\to b_1\pi)$= 8--23, 32--86 and 52--151\,MeV for $m_{1^{-+}}$ = 1.6, 1.8 and 2.0\,GeV.

\begin{table*}[hbt]
\setlength{\tabcolsep}{1.5pc}
\catcode`?=\active \def?{\kern\digitwidth}
\caption{Decay widths for $\pi_1 \to b_1\pi$.}
\label{tab:width}
\begin{tabular*}{\textwidth}{@{}l@{\extracolsep{\fill}}rrrr}
\hline
\multirow{2}{*}{}& $m_{\pi_1}$=1.6\,GeV& $m_{\pi_1}$=1.8\,GeV& $m_{\pi_1}$=2.0\,GeV\\
\cline{2-4}
&\multicolumn{3}{c}{$\Gamma(\pi_1\to b_1\pi)$/MeV}
\\
\hline
$g_{b_1}^1,g_{b_1}^2$ from $j^T$ LCSR&$>2$&$>9$&$>16$\\
\hline
$g_{b_1}^1,g_{b_1}^2$ from $j^D$ LCSR& 20--40 & 46--103 & 62--163 \\
\hline
$g_{b_1}^1$ from $j^T$ LCSR, $g_{b_1}^2$ from $j^D$ LCSR & 8--23 & 32--86 & 52--151 \\
\hline
$g_{b_1}^1$ from $j^D$ LCSR, $g_{b_1}^2$ from $j^T$ LCSR&$>12$ & $>18$ &$>22$\\
\hline
\end{tabular*}
\end{table*}

\subsection{Numerical analysis for $\pi_1\rightarrow \rho\pi$}

\begin{figure}[htbp]
\centering
\includegraphics[scale=0.62]{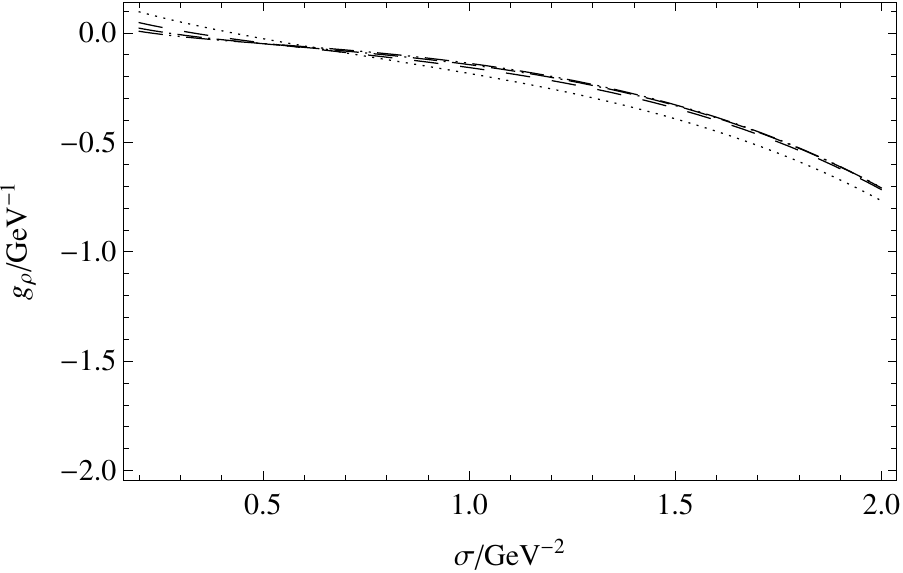}
\includegraphics[scale=0.62]{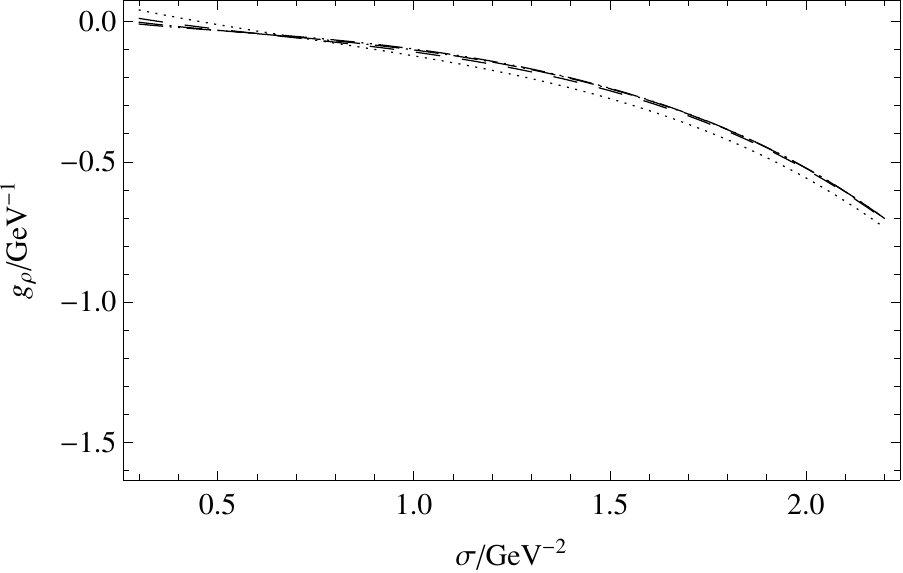}
\includegraphics[scale=0.62]{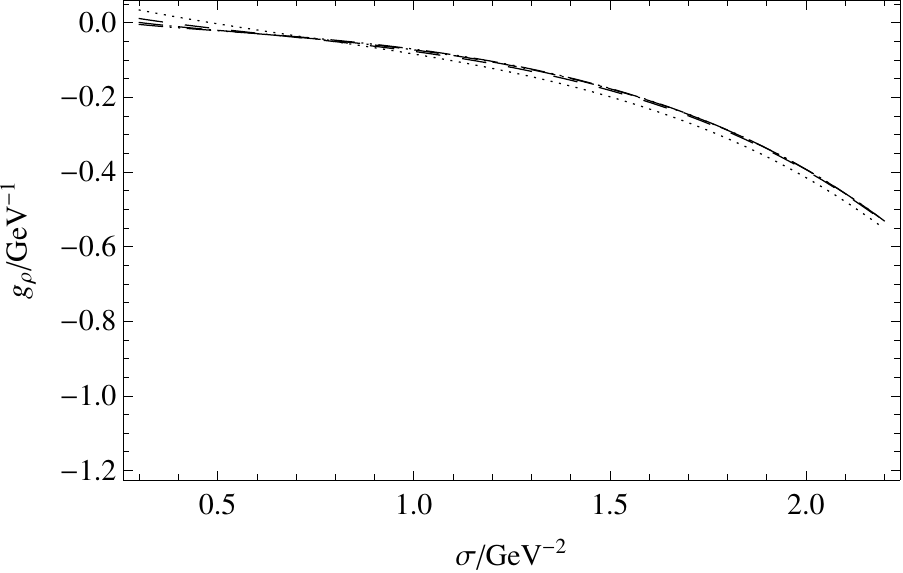}
\caption{\label{fig:grho} $g_\rho-\sigma$ curve for $m_{\pi_1}$ = 1.6\,GeV,
1.8\,GeV and 2.0\,GeV. The dotted line, the dashed line, the dot-dashed line and the dot-dot-dashed line
denote $\{s_{01}, s_{02}\}$ = \{2\,GeV$^{2}$, 3\,GeV$^2$\}, \{3\,GeV$^{2}$, 4\,GeV$^2$\},
 \{4\,GeV$^{2}$, 5\,GeV$^2$\} and \{5\,GeV$^{2}$, 6\,GeV$^2$\} respectively.
 }
\end{figure}

From the sum rules obtained from using the tensor current as the interpolating current, we can also try to obtain the prediction for $g_\rho$. In the numerical analysis we use $m_\rho=0.77\,\textrm{GeV}$ and $f^T_\rho(2~\textrm{GeV})=0.159\,\textrm{GeV}$ \cite{Jansen:2009hr,Bakulev:1999gf}. The coupling is found to be insensitive to the variation of $s_{02}$ once $s_{01}$ is fixed, thus it is reasonable to assume $s_{02}=s_{01}+1.0\,\textrm{GeV}^2$. As shown in Figure \ref{fig:grho}, although the $g_\rho$ curves do not reach exact stability in $\sigma$,  the absolute value of $g_\rho$ is small (less than 1 ${\rm GeV}^{-1})$ within the large range of $\sigma$, which suggests very small partial decay width (no more than O(MeV) on the order of magnitude). The smallness of the $\rho\pi$ decay width is easily interpreted by the light-cone expansion \eqref{eq:22}, where the leading-twist DAs are absent.

\section{Summary and conclusions}

\begin{itemize}
\item We have calculated the coefficients of the dimension-8 condensates in the OPE of the ($1^{-+}$,$0^{++}$) current--current two-point correlation function, considering the operator mixing under renormalization. We find the inclusion of these higher power corrections increases the mass predictions for the $1^{-+}$ light hybrid meson in QCD sum rules, and so does the update of the value of  the tri-gluon condensate $\langle g^3G^3\rangle$.

\item We have obtained a conservative mass range 1.72--2.60\,GeV, which only covers the mass of the $\pi_1(2015)$ and disfavors the $\pi_{1}(1600)$ and the $\pi_{1}(1400)$ to be pure hybrid states.
\item We have studied the partial decay widths of the $b_1\pi$ and $\rho\pi$ decay modes of the $1^{-+}$ light hybrid within the
 framework of light-cone QCD sum rules.

\item We have obtained the partial decay widths $\Gamma(\pi_1\to b_1\pi)$= 8--23, 32--86 and 52--151\,MeV for $m_{1^{-+}}$ = 1.6, 1.8 and 2.0\,GeV respectively which are consistent with the predictions obtained from the flux tube models \cite{Kokoski:1985isandIsgur:1985vyandPage:1998gzandSwanson:1997wy} and lattice QCD \cite{Burns:2006wz} on the order of magnitude. The relatively large partial decay widths of the $b_1\pi$ mode are compatible with the hybrid explanation of the $\pi_1(1600)$ and the $\pi_1(2015)$, both of which have been observed in the $b_1\pi$ channels.

\item We have shown in the numerical analysis the partial decay width of the $\rho\pi$ decay mode is small, which can also be deferred from the structure of the light-cone expansion of the correlation function: the leading-twist DAs are absent
    in this light-cone expansion. This result differs quite a lot from the one \cite{Huang:2010dc} obtained from using the vector current $\bar \psi\gamma_\mu \psi$ as the interpolating current in light-cone QCD sum rules. Considering the decay constant of the $\rho$ meson corresponding to the tensor current obtained from the lattice calculation \cite{Jansen:2009hr} is in a reasonable region, this discrepancy is hard to be attributed to the non-coupling of the tensor current and the $\rho$ meson. Our results also go in line with the predictions obtained from the flux tube models \cite{Kokoski:1985isandIsgur:1985vyandPage:1998gzandSwanson:1997wy,Close:1994hc}. Since there are still debates on the experimental results for the $\rho\pi$ decay mode of the $\pi_1(1600)$, and the $\pi_1(2015)$ has not yet been observed in the $\rho\pi$ final states \cite{Patrignani:2016xqp,Meyer:2015eta}, further theoretical and experimental studies of the $\rho\pi$ decay mode can be important for identifying the $1^{-+}$ light hybrid meson.
\end{itemize}

\end{document}